\documentclass{article}
\usepackage{spconf,amsmath,graphicx}
\usepackage{times}
\usepackage{epsfig}
\usepackage{graphicx}
\usepackage{amsmath}
\usepackage{amssymb}
\usepackage{xcolor}
\usepackage{subcaption}
\usepackage{adjustbox}
\usepackage{lipsum}
\usepackage{authblk}
\usepackage{cuted}

\usepackage{float}
\usepackage{makecell}

\title{LECA: A Learned Approach for Efficient Cover-agnostic Watermarking}
\name{Xiyang Luo, Michael Goebel, Elnaz Barshan, Feng Yang}
\address{Google Inc.}
\begin{document}
\maketitle
\begin{abstract}
In this work, we present an efficient multi-bit deep image watermarking method that is cover-agnostic yet also robust to geometric distortions such as translation and scaling as well as other distortions such as JPEG compression and noise. Our design consists of a light-weight watermark encoder jointly trained with a deep neural network based decoder. Such a design allows us to retain the efficiency of the encoder while fully utilizing the power of a deep neural network. Moreover, the watermark encoder is independent of the image content, allowing users to pre-generate the watermarks for further efficiency.  To offer robustness towards geometric transformations, we introduced a learned model for predicting the scale and offset of the watermarked images. Moreover, our watermark encoder is independent of the image content, making the generated watermarks universally applicable to different cover images. Experiments show that our method outperforms comparably efficient watermarking methods by a large margin.
\end{abstract}
\section{Introduction}
\label{sec:intro}

Since the late 1990s and early 2000s, there has been a rich history of research on image watermarking~\cite{voyatzis1998digital,katzenbeisser2000digital,podilchuk2001digital,hartung1999multimedia,petitcolas1999information}. Early methods, such as Least Significant Bit (LSB) embedding~\cite{bender1996techniques}, embedded a message in a cover image by modifying the LSB of the cover. This would produce minimal perturbation to the cover and provide a simple decoding scheme, but little robustness to even minor image modifications. Additional robustness was achieved for the most part through the use of spread-spectrum techniques~\cite{cox1997secure,deguillaume1999robust}. This is generally achieved by embedding messages in some transformed domain of the image, such as the Discrete Fourier Transform (DFT)~\cite{urvoy2014perceptual,deguillaume1999robust,cedillo2014robust}, Discrete Cosine Transform (DCT)~\cite{barni1998dct,liu2017digital,ernawan2018robust,hamidi2018hybrid}, or Discrete Wavelet Transform (DWT)~\cite{shi2012rst,sun2021geometrically,anand2020improved,daren2001dwt}. 

\begin{figure}[!ht]
  \includegraphics[width=0.5\textwidth]{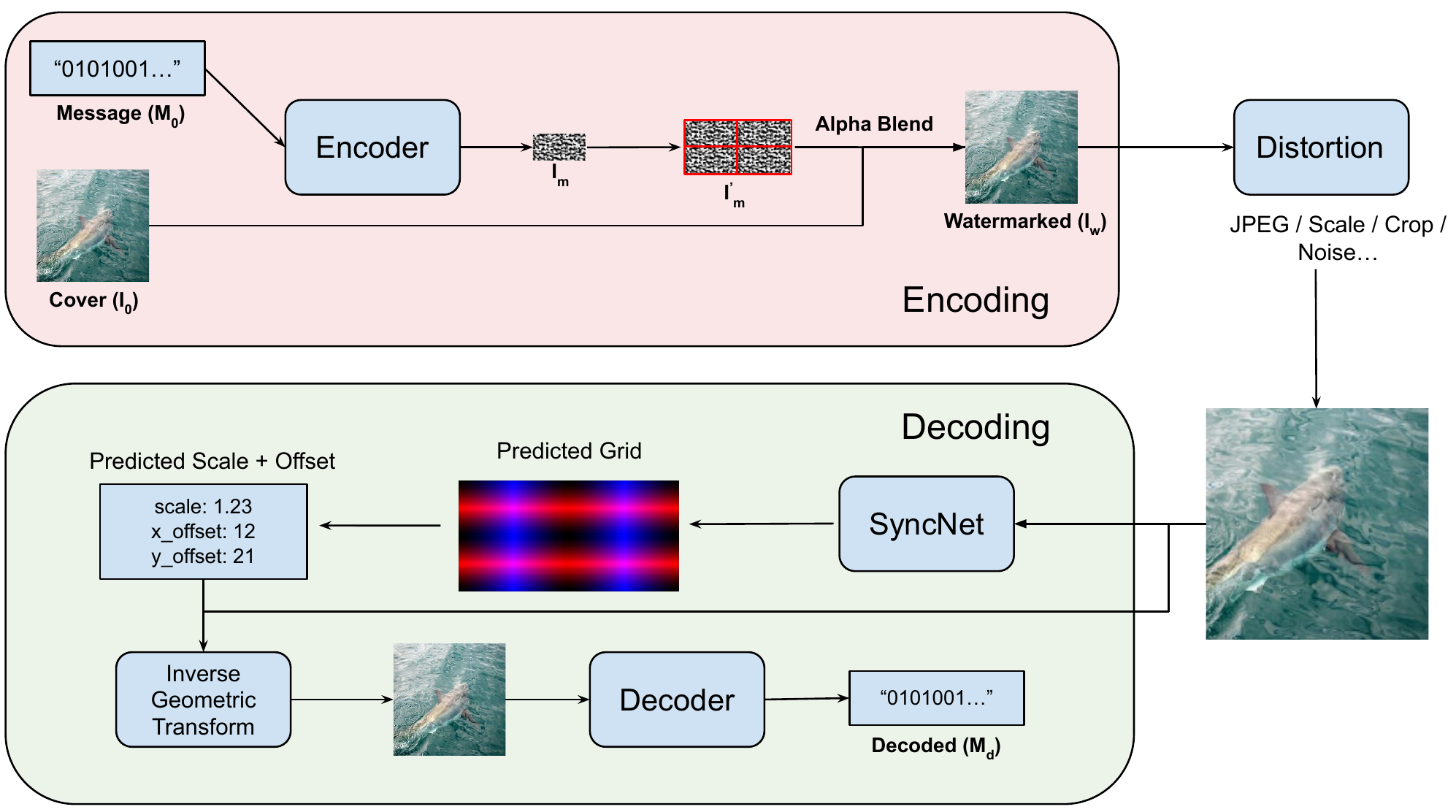}
  \centering
  \vspace{-5mm}
  \caption{Overview of the encoding and decoding pipeline of our watermarking method. The encoder network takes the input message (128 bits) and outputs a mask pattern, which is then tiled and alpha-blended with the cover to produce the watermarked image. For decoding, our method first predicts the scale and translation that have been applied to the image, inverts the transforms, and then sent to a decoder network to extract the watermark message.  }
  \label{fig:overall_design}
\end{figure} 

Particularly difficult are geometric distortions such as Rotation, Scaling, and Translation (RST). Failure to account for these will result in synchronization errors, causing a failure of the system. Several methods in classical watermarking have been developed to account for RST transformations~\cite{pereira2000robust,zheng2009rst,deguillaume2002method}. The majority of these either attempt to embed the watermark in a domain which is invariant to one or more of these transformations, or to estimate the transformation, and rectify the original image for watermark extraction. These traditional watermarking methods often have low computational cost, but are often tailored to specific distortion. More recently, deep learning based methods have made significant progress in improving the robustness and perceptual quality of image watermarking~\cite{zhu2018hidden,tancik2019stegastamp,luo2020cvpr,zhang2020udh}. However, the cost of encoding a watermark is much greater for these methods compared to traditional methods.

To this end, we propose a method that enjoys the best of both worlds by combining an light-weight encoder with a convolutional neural network (CNN) based decoding. The encoder and the decoder are jointly trained so that both components are co-optimized for common distortions such as JPEG or noise. Furthermore, we also propose a novel CNN-based solution which estimates scaling and translation to handle synchronization. Experiments show that our method outperforms comparable traditional watermarking methods by a large margin.  

Note that the encoder in our method is \emph{cover-agnostic} in the sense that it does not depend on the image content. This independence allows users to pre-generate the watermark overlay, reducing the encoding cost significantly.  Though previous works have also explored alpha-blending between the original~\cite{barni1998dct}, those methods do not utilize the power of a neural-network based decoder.  \cite{zhang2020udh} also explored a cover-agnostic watermarking through neural networks.  However, our method differs from~\cite{zhang2020udh} in the following ways. First, the computational cost of our watermark encoder is much lower compared to ~\cite{zhang2020udh}, which uses a U-net~\cite{ronneberger2015u} as backbone. We also note that~\cite{zhang2020udh} relies on an external pipeline to pre-rectify the image, which only supports the case where \emph{entire} watermarked cover is available as input. In contrast, our method supports arbitrarily cropped images. 

\section{Method}
\label{sec:method}
\subsection{Cover-agnostic Encoder}
\begin{figure}[t]
    \centering
    \begin{tabular}{c c}
    \includegraphics[ width=0.46\linewidth]{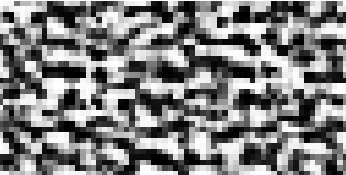} & \hspace{-3mm}
    \includegraphics[ width=0.46\linewidth]{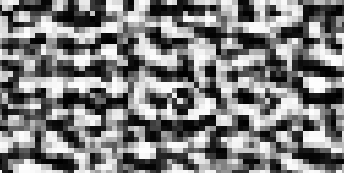}  \\ 
    \end{tabular}
    \caption{Samples of the $32\times 64$ overlay pattern for two random input messages. }
    \label{fig:overlay_sample}
    \vspace{-5mm}
\end{figure}

The watermark encoder network takes the message vector $M_0\in \{0, 1\}^m$ as input and outputs an overlay pattern of size $h \times w$. To make the encoder light-weight, we apply a single fully-connected layer to map the message space to the mask space (see Eq.~\ref{eq:encoder-net}). A sigmoid function is applied per-pixel to constrain the mask values to $[0, 1]$. The overlay image $I_m$ is then mapped to $I'_m$ by multiplying a constant color vector $c\in \mathbb{R}^3$ to each pixel, and then repeating spatially to the dimensions of the cover image (see Eq.~\ref{eq:encoder-net}). $I'_m$ is cropped from bottom-right to fit the cover image size if the cover image width and height are not multiplies of the overlay. This amounts to 
\begin{equation}
    I_m = \text{sigmoid}(W_e M_0 + b_e), \quad I'_m = \text{Repeat}(I_m \cdot c),
    \label{eq:encoder-net}
\end{equation}
where $W_e$ and $b_e$ are the weights and biases of the fully-connected layer in the encoder network.  We set the size of the mask to $h \times w = 32 \times 64$ throughout this paper, but other mask sizes are also supported depending on the application. Visualization of the patterns can be found in Figure~\ref{fig:overlay_sample}. 

The cover $I_o$ and overlay image $I'_m$ are combined through alpha-blending, as shown in Eq.~\ref{eq:encoder-alpha}, where $\alpha$ controls the strength of the watermark.
\begin{equation}
    I_w = (1 - \alpha) \times I_o + \alpha \times I'_m.
    \label{eq:encoder-alpha}
\end{equation}

\subsection{SyncNet: Predicting scaling and translation.}
Traditional watermarking methods for handling geometric transformation roughly fall into two broad categories: (1) embedding in a domain invariant to these transforms, or (2) attempting to estimate and invert the transform prior to the watermark extraction. The analogous approaches in a learning-based system would be either to directly learn invariance to these transforms in the encoder-decoder training (see Section~\ref{subsec:network-training}), or predicting the geometric transforms. We choose the latter since we empirically observe that directly learning invariance only succeeds if the geometric transforms are insignificant.

\begin{figure}[t]
    \centering
    \includegraphics[ width=0.9\linewidth]{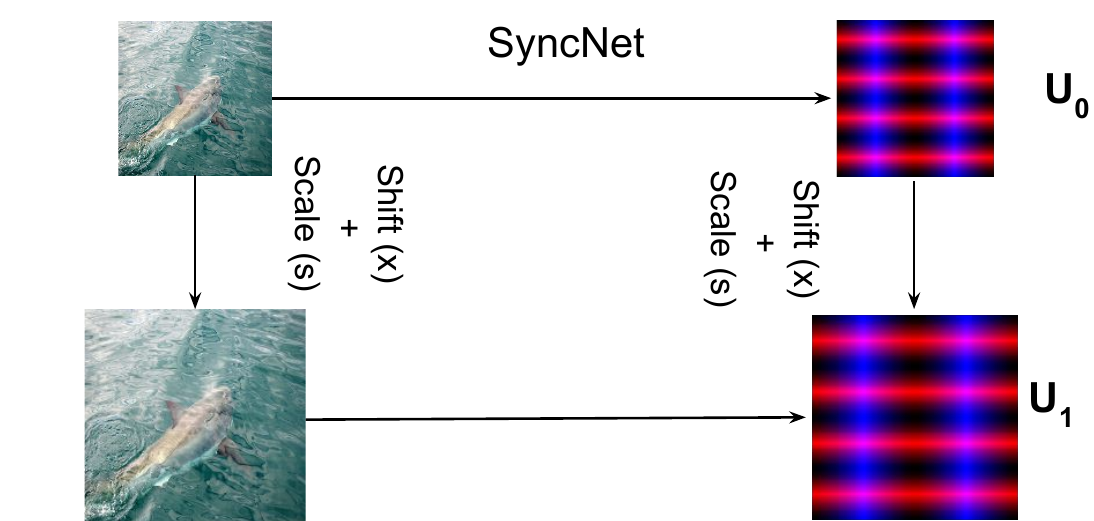} 
        \vspace{-2mm}
    \caption{Illustration of the equivariance property. The prediction is done by comparing $U_0$, a universal template pattern shared by all un-transformed images, to $U_1$, the current output.}
    \label{fig:syncnet-pattern}
    \vspace{-2mm}
\end{figure}

Directly predicting the scale and offset through a standard image classification network yields poor prediction accuracy. Instead, we present a more nuanced approach by training a neural network to learn \emph{equivariance} on the domain of images watermarked by our decoder. We name the network SyncNet since it provides a solution for translation and scaling commonly referred to as the synchronization problem.  Figure~\ref{fig:syncnet-pattern} provides an illustration of the equivariance property that the network is trained to satisfy.  We note that this property \emph{only} holds for images that have been watermarked by our system, and does not hold for arbitrary input. 

To predict the transforms that has been applied to a watermarked image, we compare $U_1$, the SyncNet output of a transformed watermarked image, with the SyncNet output of the same image pre-transform. Since we do not have access to the pre-transformed image,  we instead apply the constraint that the output of SyncNet is \emph{shared} for \emph{all} watermarked images that did not go through any geometric transformations.  We define the output  $U_0$ as the \emph{universal template}.

 A brute-force search is applied to obtain the best transform (with integer precision) from $U_0$ to $U_1$ such that $\|T(U_0) - U_1\|^2$ is minimized. For the choice of the \emph{universal template}, we select a pair of periodic signal (super-imposed in red and blue in Figure~\ref{fig:syncnet-pattern}), where the peak of the signal correspond to the $x$ (and $y$) coordinates of the \emph{\textbf{centers}} of the overlay patches after scaling and translation, \emph{e.g.}, the peaks of the horizontal lines in $U_0$ are $\{(n + \frac{1}{2})\times h, n\in \{1, 2, \dots\}\}$, where $h=32$ is the height of the overlay pattern.  Note that given the periodicity, we can only predict the offset amount \emph{modulo} the overlay patch width and height ($w=64$ and $h=32$). However, this does not affect the message decoding since the overlay patterns are tiled. 

\subsection{Decoder Network}

\begin{figure}[!ht]
    \centering
    \includegraphics[ width=1.01\linewidth]{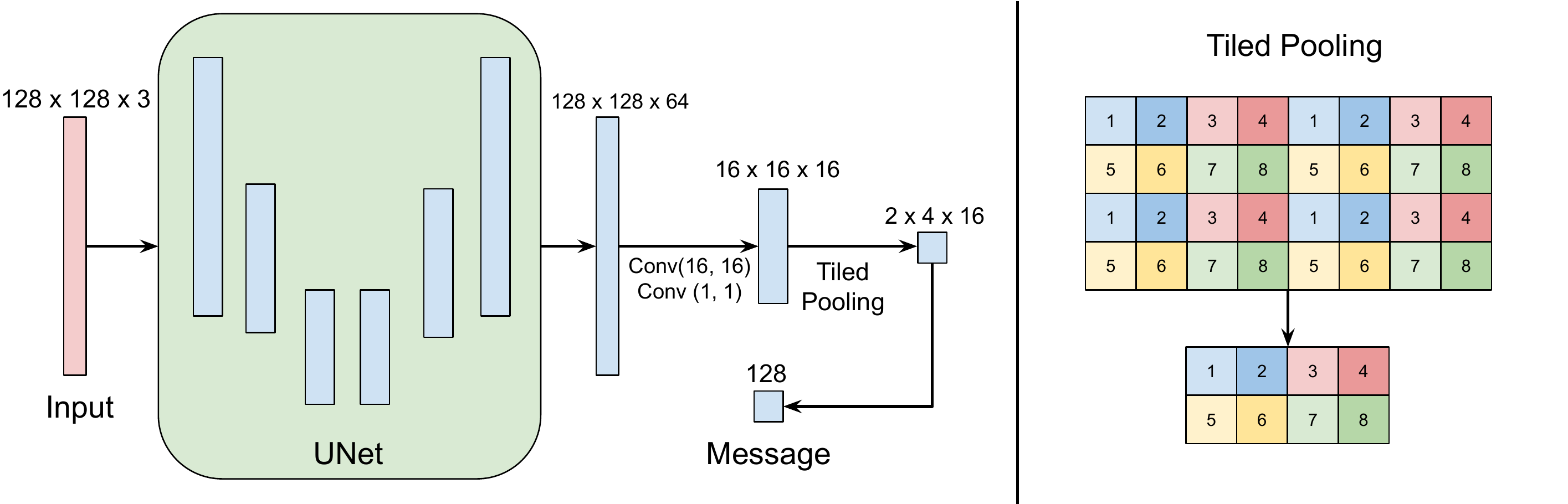}
    \caption{Architecture of the decoder network. The decoder consists of a UNet combined with a convolution layer with kernel size 16 and stride 16, and a ``tiled pooling" operation, before being reshaped. The ``tiled pooling" maps the activation of arbitrary size to a tensor of $2\times 4 \times 16$, which accounts for the spatial repetition in the overlay pattern. An illustration of the tiled pooling is on the right side of the figure. Note that we can process any image size larger than $32 \times 64$, and the dimensions in the figure are only an example for illustration.}
    \vspace {-5mm}
    \label{fig:overlay_decoder}
\end{figure}

The decoder network consists of a U-net~\cite{ronneberger2015u} backbone followed by two additional convolution layers and one pooling operation. The output is thresholded by $0.5$ to obtain the  binary message. The two convolution operations are of kernel size, stride, and output channels equal to  $(16, 16, 64)$ and $(1, 1, 16)$ respectively. The detailed designs are shown in Figure~\ref{fig:overlay_decoder}. We note that our design is fully-convolutional, and can accept any input size which is a multiple of the overlay pattern dimensions ($32 \times 64$). Inputs that are not multiples of the overlay dimension will be cropped to the nearest heights and widths before passing to the decoder.

We motivate the design of the final layers after U-net, which consists of two convolutions followed by a pooling operation. We found it beneficial to use a large kernel size for the final layer, since this increases the field-of-view of the output and can take advantage of the full size of the overlay pattern.  The ``tiled-pooling" is a simple averaging operation to account for the fact that the overlay pattern are repeated across the image. 


\subsection{Network Training}
\label{subsec:network-training}
The encoder and decoder is trained \emph{jointly} without synchronization to minimize the loss in Equation~\ref{eq:loss-encoder} below.
\begin{equation}
    \text{Loss}_{1} = \|I_w - I_o\|^2 + \text{CrossEntropy}(M_d, M_0),
    \label{eq:loss-encoder}
\end{equation}
see Figure~\ref{fig:overall_design} for notation definitions.
 To improve the robustness of the networks, we distort the watermarked image with Differentiable JPEG~\cite{shin2017jpeg} (Q=85), Gaussian noise ($\text{std}\sim\text{U}(0.01, 0.03)$), and $\text{offset} \sim \text{U}(0, 4)$ and $\text{scale}\sim\text{U}(1.0, 1.02)$. 
For SyncNet, we apply $\text{scale}\sim\text{U}(0.5, 2.0)$and $\text{offset} \sim \text{U}(-32,32)$) to the watermarked images, and minimize the $L_2$ loss between the predicted patterns and the ground truth patern. The encoder and decoder network weights are fixed during the training of SyncNet.

\vspace{-3mm}
\section{Experiments}
\label{sec:experiments}
In this section, we evaluate the performance of our method compared to methods with similar encoding complexity. Existing image watermarking methods all differ in payload, robustness, as well as dependence on the cover image. To perform a fair comparison, we choose the Robust Template Matching (RTM)~\cite{pereira2000robust} based on the following criteria: 1) Multi-bit with relatively large payload ($\geq 128 \text{bits}$); 2) Low dependence on the cover image; 3) Simultaneously robust to geometric transforms as well as other corruptions such as JPEG. The encoder and decoder of RTM consist of two distinct pieces, a template-encoding and a standard spread-spectrum differential encoding, both of which are done on the DFT domain. Both the template and the watermark are embedded in a ring of middle frequencies as a trade-off between robustness and imperceptibility. 

All methods are evaluated on a subset of the ImageNet~\cite{deng2009imagenet} validation set (using the standard train and validation split), which contains a diverse range of web images. We embed a payload of 128 bits for all experiments, and evaluate the average bit accuracy as well as the percentages of images that have bit errors less than $95\%$ (at most 6 bit errors), which can be perfectly recovered with error correction codes such as BCH codes~\cite{forney1965decoding}.

\begin{table*}[!ht]
    \begin{center}
    \begin{adjustbox}{width=0.90\textwidth}    
    \begin{tabular}{c|ccccccc}  \hline
    RTM~\cite{pereira2000robust} &  Identity & Jpg444(Q95) & Jpg444(Q90) & Jpg420(Q95) & Jpg420(Q90) &  GN(0.01) &GN(0.02)\\ \hline
    Crop(0.8) + Scale (0.75 - 1.0) & 52.01  / 4.0& 50.98 / 1.0& 50.53 / 0.0& 51.05 / 2.0& 49.99 / 0.0& 51.72 / 4.0& 51.04 / 1.0\\ \hline 
    Crop(0.8) + Scale (1.0 - 1.5) & 78.80 / 58.0& 76.73 / 52.0& 65.59 / 26.0& 77.16 / 53.0& 66.60 /29.0& 69.00 /30.0& 64.65 /18.0\\ \hline 
    Crop(0.8) + Scale (1.5 - 2.0) & 83.93 / 62.0& 83.53 / 61.0& 83.07 / 61.0& 82.56 / 59.0& 83.86 / 62.0& 74.46 /38.0& 71.80 / 30.0\\ \hline   \hline
    Ours &  Identity & Jpg444(Q95) & Jpg444(Q90) & Jpg420(Q95) & Jpg420(Q90) &  GN(0.01) &GN(0.02)\\ \hline
    Crop(0.8) + Scale (0.75 - 1.0) &  \textbf{96.4 / 90.0} & \textbf{95.62 / 89.0} & \textbf{93.44 / 83.0} & \textbf{95.17 / 90.0} & \textbf{92.87 / 84.0} & \textbf{96.46 / 92.0} & \textbf{91.72 / 81.0}\\ \hline     
    Crop(0.8) + Scale (1.0 - 1.5) &  \textbf{97.28 / 93.0}& \textbf{97.49 / 92.0} & \textbf{95.56 / 92.0}&  \textbf{94.68 / 90.0}&  \textbf{95.55 / 91.0}& \textbf{93.80 / 88.0} & \textbf{91.38 / 83.0}\\ \hline 
    Crop(0.8) + Scale (1.5 - 2.0) & \textbf{93.49 / 87.0} &  \textbf{93.40 / 86.0} &\textbf{92.65 / 85.0}&  \textbf{93.98 / 88.0}&  \textbf{95.49 / 91.0}& \textbf{92.98 / 86.0} & \textbf{86.58 / 73.0} \\ \hline   \hline  
    \end{tabular}
    \end{adjustbox}    
    \end{center}
    \caption{Decoding bit accuracy (percent) for various distortions with random scaling and cropping applied. For each cell in the table, we report both the average bit accuracy (first entry), and the percentage of images with greater than $95\%$ decoding accuracy (second entry).   The PSNR for our method is at 41dB and PSNR for RTM is 38dB.}
    \label{tab:scale-accuracy}
    \vspace{-3mm}
\end{table*}

\subsection{Image Quality}

\begin{figure}[!t]
    \centering
    \begin{tabular}{c c}
    \hspace{-3mm} \includegraphics[ width=0.48\linewidth]{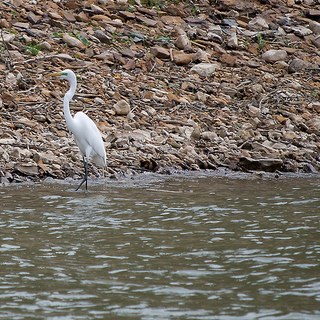} & \hspace{-3mm}
    \includegraphics[width=0.48\linewidth]{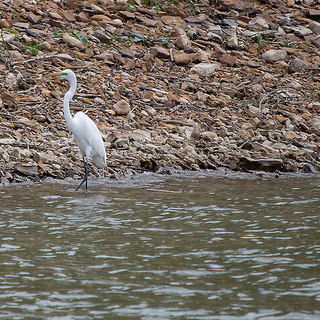}  \\ 
    \end{tabular}
    \caption{Sample of original and watermarked images. Left: Original. Right: Watermarked. }
    \label{fig:watermark_sample}
    \vspace{-3mm}
\end{figure}

We adjust the watermark strength to $\alpha=\frac{5}{255}$ which produces a nearly imperceptible watermark throughout our experiments. The fractional value is because alpha-values are usually mapped to $[0, 255]$ in browser implementations of alpha-blending. Figure~\ref{fig:watermark_sample} provides a visual sample of the watermarked image. The average peak signal-to-noise ratio (PSNR) is 40.9dB for our method. We adjust the PSNR of the compared method to match our PSNR whenever possible, and lower if the decoding accuracy is too low. 

\subsection{Decoder Robustness}
We first evaluate the standalone robustness of our decoder network compared to a standard spread-spectrum method as in RTM~\cite{pereira2000robust},~\emph{i.e.}, we do not apply any scaling or translation to the images and do not apply scale and offset predictions. Table~\ref{tab:decoder-robustness-average_accuracy} shows the average decoding accuracy evaluated on 500 images from the test set resized to $256\times 256$ with randomly generated message payload. We observe that our method outperforms the baseline by a large margin on all settings.

\begin{table}[!ht]
    \begin{center}
    \begin{adjustbox}{width=0.49\textwidth}
    \begin{tabular}{c|cccc}  \hline
             &  Identity & Jpg420(Q95) & Jpg420(Q90) &  Jpg420(Q85)\\ \hline
    RTM~\cite{pereira2000robust} & 94.94 & 92.74 & 91.30 & 85.50\\ \hline
    Ours     & \textbf{99.94} & \textbf{99.91} &  \textbf{99.88} & \textbf{99.73} \\ \hline \hline
             &  Jpg444(Q95) & Jpg444(Q90) & Jpg444(Q85) &  Jpg444(Q80)\\ \hline
    RTM~\cite{pereira2000robust} & 94.39 & 91.28 & 85.49 & 79.57\\ \hline
    Ours     &  \textbf{99.91} &  \textbf{99.88}& \textbf{99.76} & \textbf{99.58}\\ \hline \hline    
             &  GN(0.01) & GN(0.02) &  GN(0.03) & GN(0.04)\\ \hline    
    RTM~\cite{pereira2000robust} & 94.70 & 94.11 & 93.00 & 91.29\\ \hline
    Ours     &  \textbf{99.94}& \textbf{99.91}&  \textbf{99.87}& \textbf{99.75}\\ \hline
    \end{tabular}
    \end{adjustbox}
    \end{center}
    \caption{Average bit accuracy (percentage) with no scaling and translation. PSNR for both watermarking methods is adjusted to 41dB. Results are averaged across 500 images and the messages are generated at random.}
    \label{tab:decoder-robustness-average_accuracy}
    \vspace{-5mm}
\end{table}

\subsection{Decoder With Synchronization}
In this section, we evaluate the end-to-end performance of our system by adding scaling and translation on top of the existing distortions,~\emph{i.e.}, by applying ``random crop + scaling + distortion" to the watermarked images where ``distortion" is optionally JPEG or Gaussian Noise. The results are evaluated on 100 images resized to $384\times 384$. We randomly crop a square with $80\%$ width and height of the original to provide full coverage of all possible translation values, since $384\times 0.2~=77$ is already greater than the overlay patch size. Table~\ref{tab:scale-accuracy} provides a breakdown of the average decoding accuracy for three different ranges of scaling factors. Note that the decoding accuracy of our method is greater than 90\% for nearly all scenarios, and also out-performing the comparison by a large margin.

\begin{table}[!ht]
    \begin{center}
    \begin{adjustbox}{width=0.48\textwidth}    
    \begin{tabular}{c|ccccc}  \hline
         \textbf{C(0.8) + S(0.75 - 1.0) }        &  Identity &GN(0.01) &GN(0.02) &  Jpg444(Q90) &  Jpg420(Q90)\\ \hline
    Ours (Offset Error) & 3.2 & 3.5 & 5.2 & 3.6 & 4.7\\ \hline          
    Ours (Scale Error) & \textbf{0.01} & \textbf{0.006} & \textbf{0.01} & \textbf{0.008} & \textbf{0.01}\\ \hline
    RTM~\cite{pereira2000robust} (Scale Error) & 0.28 & 0.28 & 0.27 & 0.35 & 0.35\\ \hline \hline
         \textbf{C(0.8) + S(1.0 - 1.5) }        &  Identity &GN(0.01) &GN(0.02) &  Jpg444(Q90) &  Jpg420(Q90)\\ \hline    
    Ours (Offset Error) & 2.5 & 4.8 & 3.6 & 4.1 & 4.6\\ \hline          
    Ours (Scale Error) & \textbf{0.032} & \textbf{0.017} & \textbf{0.024} & \textbf{0.025} & \textbf{0.020}\\ \hline
    RTM~\cite{pereira2000robust} (Scale Error) & 0.15 & 0.18 & 0.14 & 0.23 & 0.22 \\   \hline \hline 
         \textbf{C(0.8) + S(1.0 - 1.5) }        &  Identity &GN(0.01) &GN(0.02) &  Jpg444(Q90) &  Jpg420(Q90)\\ \hline    
    Ours (Offset Error) & 3.6 & 3.8 & 4.6 & 2.9 & 2.8\\ \hline          
    Ours (Scale Error) & \textbf{0.034} & \textbf{0.035} & \textbf{0.059} & \textbf{0.030} & \textbf{0.033}\\ \hline
    RTM~\cite{pereira2000robust} (Scale Error) & 0.10 & 0.10 & 0.11 & 0.09 & 0.07 \\ 
    \hline
    \end{tabular}
    \end{adjustbox}    
    \end{center}
    \caption{Error of the scale and offset predictions for different scale factor ranges and distortions. The errors for scale and offset are presented in its absolute values,~\emph{i.e.}, an offset error of $4$ means the image is shifted by $4$ pixels.}
    \label{tab:scale-prediction-error}
    \vspace{-5mm}
\end{table}

Table~\ref{tab:scale-prediction-error} provides average error of the scale and offset predictions under the same setting as Table~\ref{tab:scale-accuracy} for our method as well as RTM. Note that since RTM acts on the magnitude of the DFT, the method is naturally translational invariant and does not require explicit translational offset prediction.

\subsection{Encoder Latency}
We measure the latency of our watermark encoder compared against RTM on images of size $384\times 384$. The average latency is \textbf{7.3ms} for our encoder compared to 29.2ms for RTM. Results are averaged across 1000 runs on a Intel(R) Xeon(R) W-2135 CPU @ 3.70GHz.

\section{Conclusion}
In this work, we present an efficient method for multi-bit image watermarking that is cover-agnostic at encoding time, and robust to scaling and translation as well as other image corruptions such as JPEG. Our solution is entirely learning-based as well as end-to-end trainable. Future directions of this work may include making the method robust to rotations.

{\small
\bibliographystyle{IEEEbib}
\bibliography{reference}

\begin{thebibliography}{10}

\bibitem{voyatzis1998digital}
George Voyatzis, Nikolaos Nikolaidis, and Ioannis Pitas,
\newblock ``Digital watermarking: An overview,''
\newblock in {\em 9th European Signal Processing Conference (EUSIPCO 1998)}.
  IEEE, 1998, pp. 1--4.

\bibitem{katzenbeisser2000digital}
S~Katzenbeisser and FAP Petitcolas,
\newblock ``Digital watermarking,''
\newblock {\em Artech House, London}, vol. 2, 2000.

\bibitem{podilchuk2001digital}
Christine~I Podilchuk and Edward~J Delp,
\newblock ``Digital watermarking: algorithms and applications,''
\newblock {\em IEEE signal processing Magazine}, vol. 18, no. 4, pp. 33--46,
  2001.

\bibitem{hartung1999multimedia}
Frank Hartung and Martin Kutter,
\newblock ``Multimedia watermarking techniques,''
\newblock {\em Proceedings of the IEEE}, vol. 87, no. 7, pp. 1079--1107, 1999.

\bibitem{petitcolas1999information}
Fabien~AP Petitcolas, Ross~J Anderson, and Markus~G Kuhn,
\newblock ``Information hiding-a survey,''
\newblock {\em Proceedings of the IEEE}, vol. 87, no. 7, pp. 1062--1078, 1999.

\bibitem{bender1996techniques}
Walter Bender, Daniel Gruhl, Norishige Morimoto, and Anthony Lu,
\newblock ``Techniques for data hiding,''
\newblock {\em IBM systems journal}, vol. 35, no. 3.4, pp. 313--336, 1996.

\bibitem{cox1997secure}
Ingemar~J Cox, Joe Kilian, F~Thomson Leighton, and Talal Shamoon,
\newblock ``Secure spread spectrum watermarking for multimedia,''
\newblock {\em IEEE transactions on image processing}, vol. 6, no. 12, pp.
  1673--1687, 1997.

\bibitem{deguillaume1999robust}
Frederic Deguillaume, Gabriela Csurka, Joseph~JK O'Ruanaidh, and Thierry Pun,
\newblock ``Robust 3d dft video watermarking,''
\newblock in {\em Security and Watermarking of Multimedia Contents}.
  International Society for Optics and Photonics, 1999, vol. 3657, pp.
  113--124.

\bibitem{urvoy2014perceptual}
Matthieu Urvoy, Dalila Goudia, and Florent Autrusseau,
\newblock ``Perceptual dft watermarking with improved detection and robustness
  to geometrical distortions,''
\newblock {\em IEEE Transactions on Information Forensics and Security}, vol.
  9, no. 7, pp. 1108--1119, 2014.

\bibitem{cedillo2014robust}
Manuel Cedillo-Hern{\'a}ndez, Francisco Garc{\'\i}a-Ugalde, Mariko
  Nakano-Miyatake, and H{\'e}ctor~Manuel P{\'e}rez-Meana,
\newblock ``Robust hybrid color image watermarking method based on dft domain
  and 2d histogram modification,''
\newblock {\em Signal, Image and Video Processing}, vol. 8, no. 1, pp. 49--63,
  2014.

\bibitem{barni1998dct}
Mauro Barni, Franco Bartolini, Vito Cappellini, and Alessandro Piva,
\newblock ``A dct-domain system for robust image watermarking,''
\newblock {\em Signal processing}, vol. 66, no. 3, pp. 357--372, 1998.

\bibitem{liu2017digital}
Shuai Liu, Zheng Pan, and Houbing Song,
\newblock ``Digital image watermarking method based on dct and fractal
  encoding,''
\newblock {\em IET image processing}, vol. 11, no. 10, pp. 815--821, 2017.

\bibitem{ernawan2018robust}
Ferda Ernawan and Muhammad~Nomani Kabir,
\newblock ``A robust image watermarking technique with an optimal
  dct-psychovisual threshold,''
\newblock {\em IEEE Access}, vol. 6, pp. 20464--20480, 2018.

\bibitem{hamidi2018hybrid}
Mohamed Hamidi, Mohamed El~Haziti, Hocine Cherifi, and Mohammed El~Hassouni,
\newblock ``Hybrid blind robust image watermarking technique based on dft-dct
  and arnold transform,''
\newblock {\em Multimedia Tools and Applications}, vol. 77, no. 20, pp.
  27181--27214, 2018.

\bibitem{shi2012rst}
Hailiang Shi, Nan Wang, Zihui Wen, Yue Wang, Huiping Zhao, and Yanmin Yang,
\newblock ``An rst invariant image watermarking scheme using dwt-svd,''
\newblock in {\em 2012 International Symposium on Instrumentation \&
  Measurement, Sensor Network and Automation (IMSNA)}. Ieee, 2012, vol.~1, pp.
  214--217.

\bibitem{sun2021geometrically}
Xue-Cheng Sun, Zhe-Ming Lu, Zhe Wang, and Yong-Liang Liu,
\newblock ``A geometrically robust multi-bit video watermarking algorithm based
  on 2-d dft,''
\newblock {\em Multimedia Tools and Applications}, vol. 80, no. 9, pp.
  13491--13511, 2021.

\bibitem{anand2020improved}
Ashima Anand and Amit~Kumar Singh,
\newblock ``An improved dwt-svd domain watermarking for medical information
  security,''
\newblock {\em Computer Communications}, vol. 152, pp. 72--80, 2020.

\bibitem{daren2001dwt}
Huang Daren, Liu Jiufen, Huang Jiwu, and Liu Hongmei,
\newblock ``A dwt-based image watermarking algorithm,''
\newblock in {\em IEEE International Conference on Multimedia and Expo, 2001.
  ICME 2001.} IEEE Computer Society, 2001, pp. 80--80.

\bibitem{pereira2000robust}
Shelby Pereira and Thierry Pun,
\newblock ``Robust template matching for affine resistant image watermarks,''
\newblock {\em IEEE transactions on image Processing}, vol. 9, no. 6, pp.
  1123--1129, 2000.

\bibitem{zheng2009rst}
Dong Zheng, Sha Wang, and Jiying Zhao,
\newblock ``Rst invariant image watermarking algorithm with mathematical
  modeling and analysis of the watermarking processes,''
\newblock {\em IEEE transactions on image processing}, vol. 18, no. 5, pp.
  1055--1068, 2009.

\bibitem{deguillaume2002method}
Fr{\'e}d{\'e}ric Deguillaume, Sviatoslav~V Voloshynovskiy, and Thierry Pun,
\newblock ``Method for the estimation and recovering from general affine
  transforms in digital watermarking applications,''
\newblock in {\em Security and watermarking of multimedia contents IV}.
  International Society for Optics and Photonics, 2002, vol. 4675, pp.
  313--322.

\bibitem{zhu2018hidden}
Jiren Zhu, Russell Kaplan, Justin Johnson, and Li~Fei-Fei,
\newblock ``Hidden: Hiding data with deep networks,''
\newblock in {\em Proceedings of the European Conference on Computer Vision
  (ECCV)}, 2018, pp. 657--672.

\bibitem{tancik2019stegastamp}
Matthew Tancik, Ben Mildenhall, and Ren Ng,
\newblock ``Stegastamp: Invisible hyperlinks in physical photographs,''
\newblock in {\em Proceedings of the IEEE/CVF Conference on Computer Vision and
  Pattern Recognition}, 2020, pp. 2117--2126.

\bibitem{luo2020cvpr}
Xiyang Luo, Ruohan Zhan, Huiwen Chang, Feng Yang, and Peyman Milanfar,
\newblock ``Distortion agnostic deep watermarking,''
\newblock in {\em Proceedings of the IEEE/CVF Conference on Computer Vision and
  Pattern Recognition (CVPR)}, June 2020.

\bibitem{zhang2020udh}
Chaoning Zhang, Philipp Benz, Adil Karjauv, Geng Sun, and In~So Kweon,
\newblock ``Udh: Universal deep hiding for steganography, watermarking, and
  light field messaging,''
\newblock {\em Advances in Neural Information Processing Systems}, vol. 33, pp.
  10223--10234, 2020.

\bibitem{ronneberger2015u}
Olaf Ronneberger, Philipp Fischer, and Thomas Brox,
\newblock ``U-net: Convolutional networks for biomedical image segmentation,''
\newblock in {\em International Conference on Medical image computing and
  computer-assisted intervention}. Springer, 2015, pp. 234--241.

\bibitem{shin2017jpeg}
Richard Shin and Dawn Song,
\newblock ``Jpeg-resistant adversarial images,''
\newblock in {\em NIPS 2017 Workshop on Machine Learning and Computer
  Security}, 2017, vol.~1.

\bibitem{deng2009imagenet}
Jia Deng, Wei Dong, Richard Socher, Li-Jia Li, Kai Li, and Li~Fei-Fei,
\newblock ``Imagenet: A large-scale hierarchical image database,''
\newblock in {\em 2009 IEEE conference on computer vision and pattern
  recognition}. Ieee, 2009, pp. 248--255.

\bibitem{forney1965decoding}
George Forney,
\newblock ``On decoding bch codes,''
\newblock {\em IEEE Transactions on information theory}, vol. 11, no. 4, pp.
  549--557, 1965.

\end{thebibliography}
}

\end{document}